# Coherent synthetic aperture imaging for visible remote sensing via reflective Fourier ptychography

MENG XIANG,[1,2] AN PAN,[1,2] YIYI ZHAO,[1] XUEWU FAN,[1] HUI ZHAO,[1,*] CHUANG LI,[1,*] BAOLI YAO[1]

[1] *Xi'an Institute of Optics and Precision Mechanics, Chinese Academy of Sciences, Xi'an 710119, China*
[2] *University of Chinese Academy of Sciences, Beijing 100049, China*
*Corresponding zhaohui@opt.ac.cn (H. Zhao) lichuang@opt.ac.cn (C. Li)

**Synthetic aperture radar (SAR) can measure the phase with antenna and microwave, which cannot be directly extended to visible light imaging due to phase lost. In this letter, we reported an active remote sensing with visible light via reflective Fourier ptychography (FP), termed coherent synthetic aperture imaging (CSAI), achieving high resolution, wide field-of-view (FOV) and phase recovery. A proof-of-concept experiment was reported with laser scanning and a collimator for the infinite object. Both smooth and rough objects are tested, and the spatial resolution increased from 15.6 $\mu m$ to 3.48 $\mu m$ with a factor of 4.5. The speckle noise can be suppressed by FP unexpectedly. Meanwhile, the CSAI method may replace the adaptive optics to tackle the aberration induced from atmospheric turbulence and optical system by one-step deconvolution.**

The spatial resolution in the traditional passive remote sensing is defined by the optical aperture with the formula of $1.22\lambda f/D$, where $\lambda$ denotes the center wavelength, $f$ is the focal length, and $D$ is the diameter of aperture [1]. The standard strategy to improve the resolution is to scale the size of the lens up from the optical design perspective, while it imposes geometric aberrations to the system that thus more optical surfaces would be required to optimize aberrations in turn. In optical lens construction, it would be very challenging with the constraint of a finite load. An active remote sensing technique, termed synthetic aperture radar (SAR), can directly measure the full complex field by the antenna with picosecond temporal resolution and obtain a higher spatial resolution with virtual aperture synthesis [2, 3]. Using visible light will have higher resolution according to the above formula, however, to make a comparable measurement using visible light, a detector would have to continuously record information with a temporal resolution higher than one femtosecond, a requirement well beyond the capabilities of modern devices. As such, current camera sensors record only the intensity of the incoming optical field, and the phase information is inevitably lost.

Fourier ptychography (FP) or Fourier ptychographic microscopy (FPM) invented in 2013 is a promising computational imaging technique [4-6], which effectively tackles the problems of phase loss, aberration-introduced artifacts, narrow depth-of-field (DOF) and the tradeoff between resolution and field-of-view (FOV) simultaneously, sharing the root with ptychography [7-10]. FP and FPM are corresponding to microscopic and macroscopic applications, respectively. Currently, it has found successful application in digital pathology and towards the development of high-precision quantitative phase imaging (QPI) [11-13], high-throughput imaging [14, 15], high-speed imaging [16], three-dimensional (3D) imaging [17,18], and biomedical applications [19, 20]. Combined with the reflective FPM [21-24], Dong et al. in 2014 [25] reported a proof-of-concept experiment toward macroscopic imaging with camera scanning, which expands the potential applied scenario of FP to active remote sensing with visible light. Compared with the microwave in SAR, the visible light would have higher resolution. And FP can also extremely increase the throughput of remote sensing and provide the extra phase information. Subsequently, Holloway et al. in 2016 and 2017 [1, 2], reported and completed a long-distance subdiffraction-limited visible imaging technique based on FP, respectively, termed SAVI, which extends the imaging distance from 0.7 to 1.5 $m$ and the distance can be designed freely according to the system parameters. However, the achievable imaging range of the SAVI system is limited according to commercial off-the-shelf products, which is similar to a finite correction system. Besides, the camera scanning scheme will change the FOV, resulting in a small overlapped FOV. And the cost of the camera array is very high.

In this letter, we expanded the FP application to active remote sensing with visible light, termed coherent synthetic aperture imaging (CSAI), to achieve aberration-free, high resolution, and large FOV complex field images with laser scanning scheme and a collimator for an infinite object. We simulated a scenario with two synchronous orbiting satellites in Fig.1(a), one of which is equipped with a camera lens and a detector, and the other satellite (cost-effective CubeSat [26]) is used to carry laser source for angle-varied illuminations. No reference beam is required in our system and only intensity images are recorded, which greatly reduces the load and experimental requirements. As a proof of concept, we simplify the simulated scenario as shown in Fig.1(b). A collimator is used to simulate the object at an infinite position. In optics, a collimator may consist of a curved mirror or lens with some type of light source or an image at its focus, which can be used to replicate a target focused at infinity with little or no parallax. The combinations of laser scanning and beam-splitting (BS) prism are used to realize reflective FP. The laser is placed at the focal plane of the illumination lens to provide plane wave for the infinite object, ignoring the energy attenuation. Therefore, our CSAI system is more in line with the actual imaging scenario of remote sensing in space, and the laser scanning scheme would be more flexible, cost-effective, and compatible with most existing satellite platforms. Note that the illumination lens, BS, collimator would disappear in practice (Fig. 1(a)). The photograph of our system is shown in Fig. 1(c). A fiber-coupled laser with the wavelength of 650 $nm$ and a bandwidth of 20 $nm$ for the sample illumination to suppress the speckle noise [18]. Then the laser is collimated and expanded by a pinhole and a convex lens with the diameter and focal length of 50.8 $mm$. The optical fiber port is fixed on a precise 3D translation stage with a step size of 1 $\mu m$. The side length of the cubic BS prism is 50.8 $mm$ (2 inches), and each transparent surface is coated with an anti-reflection coating. The focal length and F/# of the collimator is 550 $mm$ and 10, respectively, where F/# is defined by $f/D$. A commercial telephoto lens (500 $mm$ focal length, F/# from 6.3 to 50) and an 8-bit industrial camera (5.5 $\mu m$ pixel pitch, 2048 × 2048 pixels) are used in the imaging system for recording low resolution (LR) images. For each position of laser (row $m$ and column $n$) and its illumination wave vector ($k_{x,m,n}$ and $k_{y,m,n}$), the LR images are given by:

$$I_{m,n}(x,y) = \left| \mathcal{F}^{-1}\left\{ O\left(k_x - k_{x,m,n}, k_y - k_{y,m,n}\right) P\left(k_x, k_y\right) \right\} \right|^2 \quad (1)$$

where $\mathcal{F}^{-1}$ is the inverse Fourier transform, $O$ denotes the Fourier spectrum of the sample's transmission function o, $P(k_x, k_y)$ is the coherent transfer function (CTF), which acts as a low-pass filter of an imaging system, and $(k_x, k_y)$ are the two-dimensional coordinates in frequency domain with respect to $(x, y)$.

In microscopy via reflective FPM, Guo et al. [22] in 2016 added a circle of an additional LED array outside the objective lens to exceed the resolution limit of 2-fold of traditional reflective FPM [21]. However, because the system needs two groups of LED arrays, there is a sudden interval of the illumination angle. In 2016, Pacheco et al. [23] presented to use the traditional lens as the objective lens, so that the aperture diaphragm of the objective lens can be placed outside the illumination path, and only one LED array can be used. However, the scheme is not practical because it cannot use a traditional objective lens to obtain the dark field images, which is different from the modern microscopy. And Inspired by our digital spherical condenser scheme [14], Lee et al. [24] went back to the original two-group LED array scheme, and reported to use a parabolic mirror with an LED array placed at the object plane or even further away, which can achieve higher resolution with higher illumination efficiency and eliminate the interval of illumination angle between the two LED arrays. In fact, the challenge in reflective FPM is the limited working distance, so that a BS cannot be placed between the objective lens and specimen, and the system is inevitably complicated. While there is enough working distance in front of the collimator to place a BS in our system for macroscopic imaging. As a result, the system can be simplified extremely.

The performances of the CSAI method are demonstrated with a USAF resolution target in Fig.2. In our design, the F/# of the telephoto lens is set to 24 (i.e., initial 0.019 NA). The step size of laser source is set as 0.4 $mm$ and 17×17 LR images are captured, corresponding to the overlapping rate of 70% and illumination NA of 0.089 in FPM [14]. The final synthetic NA is 0.108, and it can be adjusted by changing the distance between the laser and sample. The detailed recovery procedure can be found in [4, 5]. The entire FOV of a raw intensity image with normal illumination is shown in Fig. 2(a). As shown in the close-up of the raw data (Fig. 2(a1)), the minimum feature that can be resolved is group 5, element 1, namely the resolution of 15.6 $\mu m$. With our CSAI method, the intensity and phase of reconstructions are shown in Figs. 2(a2) and 2(a3), respectively. And their close-ups are shown in Figs. 2(b2) and 2(b3). The smallest resolvable feature is improved to group 7, element 2, namely the resolution of 3.48 $\mu m$, which is 4.5 times higher than the original system. This experiment verified our prototype platform that can achieve HR, wide FOV, and phase recovery for remote sensing.

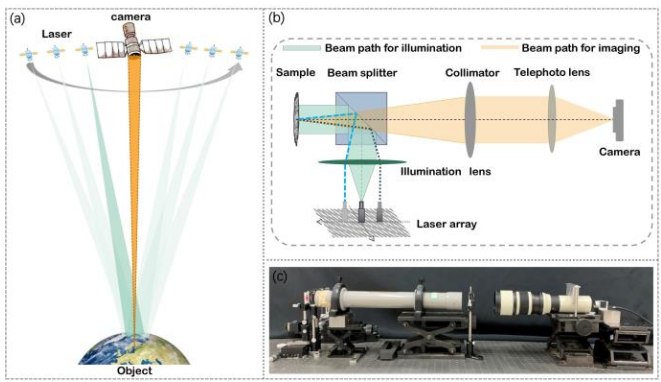

Fig. 1. Schematic and experimental photograph of the CSAI system. (a) The scenario of our CSAI system. (b) Simplified CSAI setup. (c) Photograph of our CSAI system.

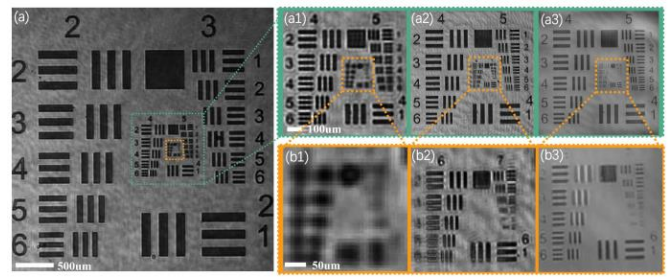

Fig. 2. Performances of the CSAI platform for the smooth object (USAF resolution target). (a) An entire FOV image captured by the CSAI platform. (a1, b1) Enlargement of the raw image. (a2, b2, a3, b3) Intensity and phase reconstructions and their close-ups, respectively, where the feature of group 7, element 2 can be clearly resolved.

In the actual remote sensing, generally, the target is optically rough surface with diffuse property, which will lead to serious speckle noise. In the past published papers in FPM, there are few studies on diffuse targets. As a demonstration, we selected a paper money of 100 CNY as the specimen. The F/# of the imaging lens is set to 50 (initial 0.0091 NA). The rest conditions are the same above. The final synthetic NA is 0.047. Figure 3(a) shows the entire FOV of LR raw image of 100 CNY. And two close-ups (green and orange regions) are shown in Figs. 3(a1) and 3(a2). The reconstructions with our CSAI system are shown in Figs. 3(b1) and 3(b2) correspondingly, while the ground truth captured by our system with aperture size of 50 *mm* (corresponding to NA 0.045) is shown in Figs. 3(c1) and 3(c2), respectively. The resolution of the Pentagram area (green region) and tower area (orange region) is significantly improved. In particular, the texture is very clear. And the speckle noise caused by the diffuse surface of the target is obviously suppressed unexpectedly, compared with Figs. 3(a1, a2) and Fig. 3(b1, b2).

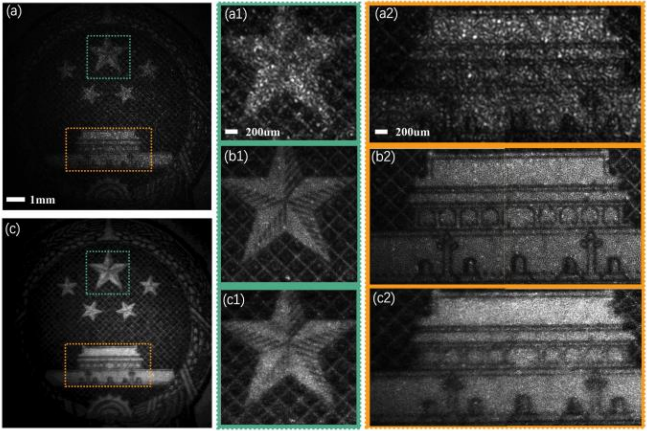

Fig. 3. Reconstructions for the rough object (100 CNY). (a)An entire FOV of raw image captured by our CSAI platform. (c) Image acquired with larger NA, which is almost equivalent to the CSAI synthetic NA of 0.045. (a1, a2) Close-ups of two different sub-regions. (b1, b2) The corresponding reconstructions. (c1, c2) The corresponding regions of ground truth.

In addition, the CTF of our system can be recovered by the FP algorithm, therefore, it may provide a method to replace the conventional adaptive optics to tackle the aberration induced from the atmospheric turbulence and optical system via one-step deconvolution [19]. In order to verify the accuracy and feasibility of the reported method, we captured an image of USAF target with incoherent illumination (Fig.4(a)), and its close-up is shown in Fig. 4(b). The recovered CTF of this region via the CSAI method is shown in Figs. 4(c) and 4(d). And the point spread function (PSF) distribution of the optical system (Fig. 4(e)) can be given by [1]:

$$PSF = \left| \mathcal{F}^{-1}\left\{CTF\right\} \right|^2 \quad (2)$$

Then the incoherent image is deconvolved by the PSF as follows [19]:

$$o(x,y) = \mathcal{F}^{-1}\left[ \frac{OTF^*(k_x,k_y)}{\left|OTF(k_x,k_y)\right|^2 + \delta} I(k_x,k_y) \right] \quad (3)$$

where $I(k_x, k_y)$ is the Fourier transform of incoherent illumination image, and $\delta$ is regularization parameters and set to 0.01 in our experiment, superscript * denotes conjugation, OTF denotes optical transfer function, which can be calculated by [1]:

$$OTF = \frac{\left|\mathcal{F}\left\{PSF\right\}\right|}{\max\left|\mathcal{F}\left\{PSF\right\}\right|} \quad (4)$$

The result of the deconvolution with the PSF result is shown in Fig. 4(g), while Fig. 4(f) provides the result of blind deconvolution as a reference. The PSF used in blind deconvolution is an ideal value calculated directly from the optical aperture. The line profiles of group 5, element 3 in Figs. 4(b-g) are shown in Figs. 4(b1-g1), respectively. Compared with blind deconvolution, the contrast in the deconvolution result with our PSF is obviously improved, which indirectly proves the correctness and feasibility of the optical aberrations recovered by the CASI platform.

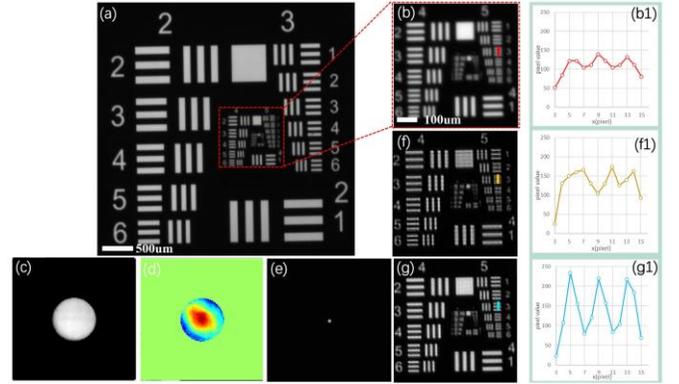

Fig. 4. Indirect verification of the aberrations recovered by our CSAI platform. (a) Entire image captured by the same system with incoherent illumination. (b) Close-up of the central region. (c-e) The amplitude and phase of recovered CTF, and the calculated PSF. (f) Blind deconvolution result. (g) The PSF deconvolution result. (b1-g1) Line profiles of group 5, element 3 in Figs. 4(b-g).

In conclusion, a low-cost CSAI platform toward active remote sensing via reflective FP is designed and verified, achieving HR, wide FOV, and phase recovery. Both smooth and rough objects are tested, and the spatial resolution is improved with the factor of 4.5. The speckle noise can be suppressed by FP unexpectedly. Meanwhile, the CSAI method may replace the adaptive optics to tackle the optical aberration. It shows the potential to be applied in remote sensing. We also notice other challenges in our system, such as laser energy attenuation, ambient stray light, target motion, and satellite orbital stability, which are expected to be the subject of future work.

**Funding.** National Natural Science Foundation of China (NSFC) (81427802 and 61975233) and Major Project on High Resolution Earth Observation System (GFZX04014307).

**Acknowledgment**. The authors thank Mingyang Yang, Zhixin Li, Yuming Wang, and Jing Wang for helpful discussion.

**References**

1. J. W. Goodman. *Introduction to Fourier Optics* (4th ed, New York, Macmillan Learning, 2017).
2. J. Holloway, M. S. Asif, M. J. Sharma, N. Matsuda, R. Horstmeyer, O. Cossairt, and A. Veeraraghavan, IEEE Trans. Comput. Imag. 2, 251-265 (2016).
3. J. Holloway, Y. Wu, M. K. Sharma, O. Cossairt, and A. Veeraraghavan, Sci. Adv. 3, e1602564 (2017).
4. G. Zheng, R. Horstmeyer, and C. Yang, Nat. Photon. **7**, 739-745 (2013).
5. X. Ou, G. Zheng, and C. Yang, Opt. Express **22**, 4960-4972 (2014).
6. A. Pan, K. Wen, and B. Yao, Opt. Lett. **44**, 2032-2035 (2019).
7. A. Pan, X. Zhang, B. Wang, Q. Zhao, and Y. Shi, Acta Phys. Sin. **65**, 014204 (2016)
8. A. Pan, D. Wang, Y. Shi, B. Yao, Z. Ma, and Y. Han, Acta Phys. Sin. **65**, 124201 (2016).
9. A. Pan, M. Zhou, Y. Zhang, J. Min, M. Lei, and B. Yao, Opt. Commun. **430**, 73-82 (2019).
10. A. Pan and B. Yao, Opt. Express **27**, 5433-5446 (2019).
11. Y. Zhang, A. Pan, M. Lei, and B. Yao, Opt. Eng. **56**, 123107 (2017).
12. A. Pan, Y. Zhang, T. Zhao, Z. Wang, D. Dan, M. Lei, and B. Yao, J. Biomed. Opt. **22**, 096005 (2017).
13. A. Pan, C. Zuo, Y. Xie, M. Lei, and B. Yao, Opt. Laser Eng. **120**, 40-48 (2019).
14. A. Pan, Y. Zhang, K. Wen, M. Zhou, J. Min, M. Lei, and B. Yao, Opt. Express **26**, 23119-23131 (2018).
15. A. C.S. Chan, J. Kim, A. Pan, H. Xu, D. Nojima, C. Hale, S. Wang, and C. Yang, Sci. Rep. **9**, 11114 (2019).
16. L. Tian, Z. Liu, L-H. Yeh, M. Chen, J. Zhong, and L. Waller, Optica **2**, 904–911 (2015).
17. R. Horstmeyer, J. Chung, X. Ou, G. Zheng, and C. Yang, Optica **3**, 827-835 (2016).
18. S. Chowdhury, M. Chen, R. Eckert, D. Ren, F. Wu, N. Repina, and L. Waller, Optica **6**, 1211-1219 (2019).
19. J. Chung, G. W. Martinez, K. C. Lencioni, S. R. Sadda, and C. Yang, Optica **6**, 647-661 (2019).
20. C. Guo, Z. Bian, S. Jiang, M. Murphy, J. Zhu, R. Wang, P. Song, X. Shao, Y. Zhang, and G. Zheng, Opt. Lett. **45**, 260-263 (2020).
21. S. Pacheco, B. Salahieh, T. Milster, J. J. Rodriguez, and R. Liang, Opt. Lett. **40**, 5343-5346 (2015).
22. K. Guo, S. Dong, and G. Zheng, IEEE J. Sel. Top. Quant. Electron. **22**, 77-88 (2016).
23. S. Pacheco, G. Zheng, and R. Liang, J. Biomed. Opt. **21**, 026010 (2016).
24. H. Lee, B. H. Chon, and H. K. Ahn, Opt. Express **27**, 34382-34391 (2019).
25. S. Dong, R. Horstmeyer, R. Shiradkar, K. Guo, X. Ou, Z. Bian, H. Xin, and G. Zheng, Opt. Express **22**, 13586–13599 (2014).
26. C. Underwood, S. Pellegrino, V. J. Lappas, C. P. Bridges, and J. Baker, Acta Astronaut. **114**, 112-122 (2015).

## Full References

1. J. W. Goodman. *Introduction to Fourier Optics* (4th ed, New York, Macmillan Learning, 2017).
2. J. Holloway, M. S. Asif, M. J. Sharma, N. Matsuda, R. Horstmeyer, O. Cossairt, and A. Veeraraghavan, "Toward long-distance subdiffraction imaging using coherent camera arrays," IEEE Trans. Comput. Imag. **2**(3), 251-265 (2016).
3. J. Holloway, Y. Wu, M. K. Sharma, O. Cossairt, and A. Veeraraghavan, "SAVI: Synthetic apertures for long-range, subdiffraction-limited visible imaging using Fourier ptychography," Sci. Adv. **3**(4), e1602564 (2017).
4. G. Zheng, R. Horstmeyer, and C. Yang, "Wide-field, high-resolution Fourier ptychographic microscopy," Nat. Photon. **7**(9), 739-745 (2013).
5. X. Ou, G. Zheng, and C. Yang, "Embedded pupil function recovery for Fourier ptychographic microscopy," Opt. Express **22**(5), 4960-4972 (2014).
6. A. Pan, K. Wen, and B. Yao, "Linear space-variant optical cryptosystem via Fourier ptychography," Opt. Lett. **44**(8), 2032-2035 (2019).
7. A. Pan, X. Zhang, B. Wang, Q. Zhao, and Y. Shi, "Experimental study on three-dimensional ptychography for thick sample," Acta Phys. Sin. **65**(1), 014204 (2016)
8. A. Pan, D. Wang, Y. Shi, B. Yao, Z. Ma, and Y. Han, "Incoherent ptychography in Fresnel domain with simultaneous multi-wavelength illumination," Acta Phys. Sin. **65**(12), 124201 (2016).
9. A. Pan, M. Zhou, Y. Zhang, J. Min, M. Lei, and B. Yao, "Adaptive-window angular spectrum algorithm for near-field ptychography," Opt. Commun. **430**, 73-82 (2019).
10. A. Pan and B. Yao, "Three-dimensional space optimization for near-field ptychography," Opt. Express **27**(4), 5433-5446 (2019).
11. Y. Zhang, A. Pan, M. Lei, and B. Yao, "Data preprocessing methods for robust Fourier ptychographic microscopy," Opt. Eng. **56**(12), 123107 (2017).
12. A. Pan, Y. Zhang, T. Zhao, Z. Wang, D. Dan, M. Lei, and B. Yao, "System calibration method for Fourier ptychographic microscopy," J. Biomed. Opt. **22**(9), 096005 (2017).
13. A. Pan, C. Zuo, Y. Xie, M. Lei, and B. Yao, "Vignetting effect in Fourier ptychographic microscopy," Opt. Laser Eng. **120**, 40-48 (2019).
14. A. Pan, Y. Zhang, K. Wen, M. Zhou, J. Min, M. Lei, and B. Yao, "Subwavelength resolution Fourier ptychography with hemispherical digital condensers," Opt. Express **26**(18), 23119-23131 (2018).
15. A. C.S. Chan, J. Kim, A. Pan, H. Xu, D. Nojima, C. Hale, S. Wang, and C. Yang, "Parallel Fourier ptychographic microscopy for high-throughput screening with 96 cameras (96 Eyes)," Sci. Rep. **9**, 11114 (2019).
16. L. Tian, Z. Liu, L-H. Yeh, M. Chen, J. Zhong, and L. Waller, "Computational illumination for high-speed in vitro Fourier ptychographic microscopy," Optica **2**(10), 904–911 (2015).
17. R. Horstmeyer, J. Chung, X. Ou, G. Zheng, and C. Yang, "Diffraction tomography with Fourier ptychography," Optica **3**(8), 827-835 (2016).
18. S. Chowdhury, M. Chen, R. Eckert, D. Ren, F. Wu, N. Repina, and L. Waller, "High-resolution 3D refractive index microscopy of multiple-scattering samples from intensity images," Optica **6**(9), 1211-1219 (2019).
19. J. Chung, G. W. Martinez, K. C. Lencioni, S. R. Sadda, and C. Yang, "Computational aberration compensation by coded-aperture-based correction of aberration obtained from optical Fourier coding and blur estimation," Optica **6**(5), 647-661 (2019).
20. C. Guo, Z. Bian, S. Jiang, M. Murphy, J. Zhu, R. Wang, P. Song, X. Shao, Y. Zhang, and G. Zheng, "OpenWSI: a low-cost, high-throughput whole slide imaging system via single-frame autofocusing and open-source hardware," Opt. Lett. **45**(1), 260-263 (2020).
21. S. Pacheco, B. Salahieh, T. Milster, J. J. Rodriguez, and R. Liang, "Transfer function analysis in epi-illumination Fourier ptychography," Opt. Lett. **40**(22), 5343-5346 (2015).
22. K. Guo, S. Dong, and G. Zheng, "Fourier ptychography for brightfield, darkfield, reflective, multi-slice, and fluorescence imaging," IEEE J. Sel. Top. Quant. Electron. **22**(4), 77-88 (2016).
23. S. Pacheco, G. Zheng, and R. Liang, "Reflective Fourier ptychography," J. Biomed. Opt. **21**(2), 026010 (2016).
24. H. Lee, B. H. Chon, and H. K. Ahn, "Reflective Fourier ptychographic microscopy using a parabolic mirror," Opt. Express **27**(23), 34382-34391 (2019).
25. S. Dong, R. Horstmeyer, R. Shiradkar, K. Guo, X. Ou, Z. Bian, H. Xin, and G. Zheng, "Aperture-scanning Fourier ptychography for 3D refocusing and super-resolution macroscopic imaging," Opt. Express **22**(11), 13586–13599 (2014).
26. C. Underwood, S. Pellegrino, V. J. Lappas, C. P. Bridges, and J. Baker, "Using CubeSat/micro-satellite technology to demonstrate the Autonomous Assembly of a Reconfigurable Space Telescope (AAReST)," Acta Astronaut. **114**, 112-122 (2015).